\documentclass[a4paper,12pt]{article}

\usepackage[dvips]{graphics}
\usepackage[mathscr]{eucal}    % \mathscr
\usepackage{amsmath} % \mathfrak
\begin{document}
\begin{titlepage}
\begin{flushright}
hep-th/0508061\\
TIT/HEP-540\\
August, 2005\\
\end{flushright}
\vspace{0.5cm}
\begin{center}
{\Large \bf 
Quantum Corrections in 2D SUSY $CP^{N-1}$ Sigma Model on Noncommutative Superspace
}
\lineskip .45em
\vskip1.5cm
{\large $^\dagger$Kazutoshi Araki\footnote{E-mail: araki@phys.chuo-u.ac.jp}, 
$^\dagger$Takeo Inami\footnote{E-mail: inami@phys.chuo-u.ac.jp}, \\
$^\ddagger$Hiroaki Nakajima\footnote{E-mail: nakajima@th.phys.titech.ac.jp} 
and $^\dagger$Yorinori Saito}
\vskip 1.5em
{\large\itshape $^\dagger$Department of Physics, Chuo University\\
\itshape Kasuga, Bunkyo-ku, Tokyo, 112-8551, Japan.

\itshape $^\ddagger$Department of Physics, Tokyo Institute of 
Technology \\
\itshape O-okayama, Meguro-ku, Tokyo, 152-8551, Japan. }  \vskip 4.5em
\end{center}
%\vskip1cm

\begin{abstract}
We investigate quantum corrections in two-dimensional $CP^{N-1}$ supersymmetric nonlinear sigma model on noncommutative superspace. 
We show that this model is renormalizable, the $\mathcal{N}$=2 SUSY sector is not affected by the C-deformation 
and that the non(anti)\-commutativity parameter $C^{\alpha\beta}$ receives infinite renormalization at one-loop order. And it is the renormalizability of the model at one-loop order.
\end{abstract}
\end{titlepage}

\newpage

\section{Introduction}
One recent development in quantum field theories is the emergence of new classes of field theories with some novel properrties from the study of string theory. String theory with some of its field components having classical values on D-brane gives rise to new field theories in its low energy limits. In particular, $\mathcal{N}$=$1$ noncommutative (NC) superspace considered some time ago \cite{d-susy} can be derived in this way from superstring theory with self-dual graviphoton background values \cite{ov,bgn,seiberg,bs}. 

In four dimensions, the non(anti)commutativity is introduced to $\mathcal{N}$=$1$ superspace by deforming the anticommuting relations for superspace coordinates $\theta^{\alpha}$, $\bar{\theta}^{\dot{\alpha}}$ as \cite{seiberg}
\begin{eqnarray}
\{\theta^{\alpha} , \theta^{\beta}\}=C^{\alpha\beta},
\quad \{\theta^{\alpha} , \bar{\theta}^{\dot{\alpha}}\}=\{\bar{\theta}^{\dot{\alpha}} , \bar{\theta}^{\dot{\beta}}\}=0 .
\end{eqnarray}
The noncommutativity (NC) parameter $C^{\alpha \beta}$ is related to the (constant) graviphton background value. Due to the non-zero anticommutator of $\theta^{\alpha}$, the field theory on the superspace keeps only a half of supersymmetry (SUSY), $\mathcal{N}$=$1/2$ SUSY, some times also called deformed SUSY. Because $\theta$ and $\bar{\theta}$ are treated independently, the field theory on $\mathcal{N}$=$1/2$ superspace can be defined only in Eucledian space and such field theories lack Hermiticity. 

$\mathcal{N}$=$1/2$ SUSY field theories can be dealt with straightforwardly by modifying the product of superfields.
The product is Weyl-ordered in $\theta$ by using the Moyal product, which is defined by 
\begin{eqnarray}
f \ast g = f \exp ( -\frac{1}{2}C^{\alpha\beta}\overleftarrow{Q}_{\alpha}
                          \overrightarrow{Q}_{\beta} ) g ,
\end{eqnarray}
where $Q_{\alpha}$ is the supercharge.

%	\item Other models on NC superspace.\\
The Lagrangian of $\mathcal{N}$=$1/2$ SUSY field theories consists of two parts, the orginal $\mathcal{N}$=$1$ SUSY Lagrangian and the new terms containing $C^{\alpha\beta}$ as a coupling constant. We refer to the two parts as the non-deformed SUSY sector and the C-deformation sector, respectively. The immediate question on $\mathcal{N}$=$1/2$ SUSY field theories in perturbation is whether they are renormalizable, despite the fact that the new coupling constant $C^{\alpha \beta}$ has a negative dimension (in mass). One may further ask whether good ultraviolet (UV) properties of field theories with (extended) SUSY are preserved after adding C-deformed terms. The first question has been answered positively for
$\mathcal{N}$=$1/2$ Wess-Zumino model \cite{Grisaru:2003fd,Britto:2003kg,Romagnoni:2003xt,Berenstein:2003sr} and $\mathcal{N}$=$1/2$ supersymmetric gauge theory \cite{Berenstein:2003sr,Alishahiha:2003kg,Jack:2004pq,Araki:2003se,Terashima:2003ri}. The NC parameter $C^{\alpha\beta}$ has been shown to receive infinite renormaliation. 
Quantum corrections to %$\mathcal{N}$=1 
the non-deformed SUSY sector of these models are not affected by the C-deformation \cite{Alishahiha:2003kg,Jack:2004pq}.  

%	\item Super Y-M and Super NLSM.\\
	$\mathcal{N}$=$2$ $CP^{N-1}$ supersymmetric nonlinear sigma models (SNLSM) \cite{D'Adda:1978kp,Cremmer:1978bh} are a low-energy effective theory of 
four-dimensional $\mathcal{N}$=$1$ super Yang-Mills theories. 
Two-dimensional $\mathcal{N}$=$2$ and $\mathcal{N}$=$4$ SNLSM have remarkably good UV properties beyond perturbative renormalizability \cite{Grisaru:1986px,Alvarez-Gaume:1981hn,Alvarez-Gaume:1981hm,Morozov:1984ad}.
In non-perturbative aspects, two-dimensional SNLSM have instantons, like super Yang-Mills theories.

%	\item Purpose.\\
	Since the K\"ahler potential of SNLSM is generally non-polynomial, the action of SNLSM on noncommutative superspace 
has infinitely many terms \cite{nlsm1,abbp,agvm,Chandrasekhar:2004ti}. It is difficult to study the properties of such models even in perturbation. Fortunately the action takes a simple closed form in the case of $CP^{N-1}$ SNLSM on NC superspace using the K\"ahler quotient \cite{Ina-Naka}.

The purpose of the present paper is to investigate one-loop quantum corrections in the 
two-dimensional $CP^{N-1}$ SNLSM on NC superspace. In particular, we are interested in the loop effects on the non-deformed SUSY sector due to the new coupling $C^{\alpha\beta}$ and vice versa.
We have found that the C-deformation term of this model receives a divergent correction 
which is absorbed by the renormalization of the coupling constant $\lambda$ and the NC parameter $C^{\alpha\beta}$.

\section{2D $CP^{N-1}$ SNLSM on NC superspace}
$CP^{N-1}$ SNLSM in $d=2$ can be obtained
from that in $d=4$ by dimensional reduction \cite{Aoyama:1980yw}. The same method can be used to obtain the $CP^{N-1}$ SNLSM on NC superspace in $d=2$ \cite{Ina-Naka}.

We denote the scalar and fermion fields after solving the $CP^{N-1}$ constraints by $\varphi^{a}$, $\bar{\varphi}^{\bar{a}}$ and $\chi^{a}$, $\bar{\chi}^{\bar{a}}$ $(a=1,2,\cdots,N-1)$. The Lagrangian is written in terms of the component fields as
\begin{eqnarray}
\mathcal{L} &=& \mathcal{L}_0 + \mathcal{L}_C , \\
\mathcal{L}_0 &=& \frac{1}{\lambda} g_{a\bar{b}}\partial_{\mu}\varphi^{a}\partial^{\mu}\bar{\varphi}^{\bar{b}} + \frac{1}{\lambda} ig_{a\bar{b}}\bar{\chi}^{\bar{b}}_{+} \gamma^{\mu} D_{\mu}\chi^{a}_{+}
+ \frac{1}{\lambda} ig_{a\bar{b}}\bar{\chi}^{\bar{b}}_{-} \gamma^{\mu} D_{\mu} \chi^{a}_{-} \nonumber \\
& &  + \frac{1}{4\lambda} R_{a\bar{b}c\bar{d}}(\bar{\chi}^{\bar{b}}_{+} \gamma^{\mu} \chi^{a}_{+} )(\bar{\chi}^{\bar{d}}_{-} \gamma_{\mu} \chi^{c}_{-} ) , \label{L0}\\
\mathcal{L}_C &=& \frac{1}{\lambda} g_{a\bar{b}} g_{c\bar{d}} ( C^{11} \chi^{a}_{+} \chi^{c}_{+} - C^{22} \chi^{a}_{-} \chi^{c}_{-} ) \epsilon^{\mu \nu} (\partial_{\mu}\bar{\varphi}^{\bar{b}})(\partial_{\nu}\bar{\varphi}^{\bar{d}}) .\label{LC}
\end{eqnarray}
Here $\mathcal{L}_0$ is the non-deformed part, namely, the usual SUSY 
$CP^{N-1}$ Lagrangian \cite{D'Adda:1978kp,Cremmer:1978bh}, $\mathcal{L}_C$ is the new term due to superspace
noncommutativity.
$\chi^{a}_{+},\chi^{a}_{-}$ are the two components of the 2D spinor $\chi^{a}$. $g_{a\bar{b}}$ is the Fubini-Study metric on $CP^{N-1}$. $\Gamma^{a}_{bc}$ and $R_{a\bar{b}c\bar{d}}$ are the Christoffel symbol and the Riemann curvature tensor, respectively. $D_{\mu}\chi^{a}_{\pm}$ is the covariant derivative. They are given by
\begin{eqnarray}
g_{a\bar{b}} &=& \frac{(1 + \bar{\varphi} \varphi ) \delta_{a\bar{b}} - \bar{\varphi}_{a} \varphi_{\bar{b}}}{(1 + \bar{\varphi}\varphi)^2} ,\\
\Gamma^{a}_{bc} &=& g^{a\bar{d}}\partial_{b}g_{c\bar{d}},\\ 
D_{\mu}\chi^{a}_{\pm} &=& \partial_{\mu}\chi^{a}_{\pm} + 
 \Gamma^{a}_{bc}(\partial_{\mu}\varphi^{b})\chi^{c}_{\pm},\\
R_{a\bar{b}c\bar{d}} &=& -g_{a\bar{e}}\partial_{c}(g^{f\bar{e}}\partial_{\bar{d}}g_{f\bar{b}})=g_{a\bar{b}}g_{c\bar{d}}+g_{a\bar{d}}g_{c\bar{b}}.
\end{eqnarray}

  \subsection{Background field method and K\"ahler normal coodinates}
We evaluate the quantum corrections in our model in perturbation
in the coupling constants $\lambda$ and $C^{\alpha \beta}$.
This can be made by means of the background field method. The
familiar Riemann normal coordinates are not good for the present purpose, because they
are not holomorphic on a K\"ahler manifold. Instead we should use the expansion in terms of 
K\"ahler normal coordinates \cite{Higashijima:2000wz}, which are holomorphic coordinates. 
This expansion provides a manifestly covariant background field method preserving the complex structure on $CP^{N-1}$.
We denote the background field by $\varphi^a$, the quantum field by $\varphi_{S}^a$, 
and the holomorphic coordinates on the $CP^{N-1}$ by $\xi^a$. 
$\varphi^a$, $\partial_{\mu} \varphi^{a}$, $g_{a\bar{b}}$, and $R_{a\bar{b}c\bar{d}}$ are written in terms of K\"ahler normal coordinates as
\begin{eqnarray}
\varphi^a & \to & \varphi^a + \varphi_{S}^a (\sqrt[]{\mathstrut \lambda}\xi) ,  \\
\partial_{\mu} \varphi^{a}_{S}(\sqrt[]{\mathstrut \lambda}\xi) & = & \sqrt[]{\mathstrut \lambda} D_{\mu} \xi^{a} - \lambda \frac{1}{2} \partial^{\mu} \bar{\varphi}^{\bar{b}} R^{a}_{c\bar{b}d} \xi^{c} \xi^{d}  + O(\xi^{3}) ,\\
g_{a\bar{b}}(\varphi + \varphi_{S}(\sqrt[]{\mathstrut \lambda}\xi) ) & = & g_{a\bar{b}} + \lambda R_{a\bar{b}c\bar{d}} \xi^{c} \bar{\xi}^{\bar{d}}  + O(\xi^{3}) , \\
R_{a\bar{b}c\bar{d}}(\varphi + \varphi_{S}(\sqrt[]{\mathstrut \lambda}\xi)) & = & R_{a\bar{b}c\bar{d}} + \lambda g^{e\bar{f}} ( R_{e(\bar{b}g\bar{d}} R_{a\bar{h}c)\bar{f}} - R_{e\bar{b}g\bar{d}} R_{a\bar{h}c\bar{f}} ) \xi^{g} \bar{\xi}^{\bar{h}}  \nonumber \\ &&  + O(\xi^{3}) .
\end{eqnarray}
%Propagator
%\begin{eqnarray*}
%<\xi^{a} (x) \bar{\xi}^{\bar{b}} (y) > & = & g^{a\bar{b}} \Delta_{F}(x-y)
%\end{eqnarray*}

In the following computation of one-loop effects, we need the terms of second order in the fluctuations $\xi^a$ of the Lagrangian \eqref{L0}, \eqref{LC}. They are given by
\begin{eqnarray}
\mathcal{L}_0^{(2)} &=& g_{a\bar{b}} D_{\mu} \xi^{a} D^{\mu} \bar{\xi}^{\bar{b}} \notag \\
& & + R_{a\bar{b}c\bar{d}} ( \partial_{\mu}\varphi^{a}\partial^{\mu}\bar{\varphi}^{\bar{b}} \xi^{c} \bar{\xi}^{\bar{d}} - \frac{1}{2} \partial_{\mu} \bar{\varphi}^{\bar{b}} \partial^{\mu} \bar{\varphi}^{\bar{d}} \xi^{a} \xi^{c} - \frac{1}{2} \partial_{\mu} \varphi^{a} \partial^{\mu} \varphi^{c} \bar{\xi}^{\bar{b}} \bar{\xi}^{\bar{d}} )  \nonumber \\
& & + iR_{a\bar{b}c\bar{d}} ( \bar{\chi}^{\bar{b}}_{+} \gamma^{\mu} D_{\mu} \chi^{a}_{+} \xi^{c} \bar{\xi}^{\bar{d}} + \bar{\chi}^{\bar{b}}_{+} \gamma^{\mu} \chi^{c}_{+} D_{\mu} \xi^{a} \bar{\xi}^{\bar{d}} ) \nonumber \\
& & + iR_{a\bar{b}c\bar{d}} ( \bar{\chi}^{\bar{b}}_{-} \gamma^{\mu} D_{\mu} \chi^{a}_{-} \xi^{c} \bar{\xi}^{\bar{d}} + \bar{\chi}^{\bar{b}}_{-} \gamma^{\mu} \chi^{c}_{-} D_{\mu} \xi^{a} \bar{\xi}^{\bar{d}} ) \label{L0(2)} \nonumber \\
& & + g^{e\bar{f}} ( R_{e(\bar{b}g\bar{d}} R_{a\bar{h}c)\bar{f}} - R_{e\bar{b}g\bar{d}} R_{a\bar{h}c\bar{f}} ) \xi^{g} \bar{\xi}^{\bar{h}} ( \bar{\chi}^{\bar{b}}_{+} \gamma^{\mu} \chi^{a}_{+} )( \bar{\chi}^{\bar{d}}_{-} \gamma_{\mu} \chi^{c}_{-} )  ,
\end{eqnarray}
\begin{eqnarray}
\mathcal{L}_C^{(2)} &=& \{ g_{a\bar{b}} g_{c\bar{d}} ( D_{\mu} \bar{\xi}^{\bar{b}} D_{\nu} \bar{\xi}^{\bar{d}} - \frac{1}{2} \partial_{\mu} \varphi^{e} \partial_{\nu}\bar{\varphi}^{\bar{d}} R^{\bar{b}}_{\bar{f}e\bar{g}} \bar{\xi}^{\bar{f}} \bar{\xi}^{\bar{g}} \notag \\
& & - \frac{1}{2} \partial_{\nu} \varphi^{e} \partial_{\mu}\bar{\varphi}^{\bar{b}} R^{\bar{d}}_{\bar{f}e\bar{g}} \bar{\xi}^{\bar{f}} \bar{\xi}^{\bar{g}} ) + g_{a\bar{b}} \partial_{\mu}\bar{\varphi}^{\bar{b}} \partial_{\nu}\bar{\varphi}^{\bar{d}} R_{c\bar{d}e\bar{f}} \xi^{e} \bar{\xi}^{\bar{f}} \notag \\
& & + g_{c\bar{d}} \partial_{\mu}\bar{\varphi}^{\bar{b}} \partial_{\nu}\bar{\varphi}^{\bar{d}} R_{a\bar{b}e\bar{f}} \xi^{e} \bar{\xi}^{\bar{f}} \} \epsilon^{\mu \nu} ( C^{11} \chi^{a}_{+} \chi^{c}_{+} - C^{22} \chi^{a}_{-} \chi^{c}_{-} ) \label{LC(2)} .
\end{eqnarray}

\section{One-loop perturbative corrections}
We begin by noting that all terms in the C-deformation part
$\mathcal{L}_C$ contain  $\chi^a \chi^b$ and are non-Hermitian. In
perturbation, only the combination $\chi^a \bar{\chi}^b$
is contracted in the Wick's theorem; the combination $\chi^a \chi^b$
is not contracted.
Hence the C-deformed terms with $\chi^a \chi^b$ cannot appear as a part
of internal lines. They can only appear as external
lines. This means that there appear no loop corrections due to the new term $\mathcal{L}_C$
 to the non-deformed part of the effective action. 
The same property holds for super Yang-Mills theories \cite{Alishahiha:2003kg,Jack:2004pq}.

    \subsection{Non-deformed sector} 
Because of the absence of loop corrections due to $\mathcal{L}_{C}$, the loop corrections to the non-deformed SUSY sector in our model are the same as those in the ordinary model. In particuler, the $\beta$-function of $\lambda$ is not affected by C-deformation. For later use we summarize the previous result on the renormalization of the non-deformed SUSY sector \cite{Alvarez-Gaume:1981hn,Alvarez-Gaume:1981hm}. We use the dimensional regularization, setting $\epsilon = 1-d/2$, where $d$ is the space-time dimension.
The divergent one-loop correction to the non-deformed sector is written as that to $\mathcal{L}_{0}$. 
\begin{eqnarray}
\delta \mathcal{L}_{{\rm Boson}} & = & R_{a\bar{b}c\bar{d}}\partial_{\mu}\varphi^{a}\partial^{\mu}\bar{\varphi}^{\bar{b}} 
  \dot{\xi}^{c}  \dot{\bar{\xi}}^{\bar{d}}   
\nonumber \\
                    & = &  \frac{1}{2 \epsilon} \frac{1}{2\pi} R_{a\bar{b}}\partial_{\mu}\varphi^{a}\partial^{\mu}\bar{\varphi}^{\bar{b}} \nonumber \\
                    & = &  \frac{1}{2 \epsilon} \frac{\lambda N}{2\pi} \mathcal{L}_{{\rm Boson}}.
\end{eqnarray}
Here $\cdot$ (or $\cdot\cdot$) over a pair of fields $\xi^a (x)$ mean that they are contracted. We have used the relation which holds for $CP^{N-1}$, 
\begin{equation}
R_{a\bar{b}} = N g_{a\bar{b}}.
\end{equation}

The bare quantities will be denoted by the suffix 0. $g_{a\bar{b}}/\lambda_{0}$  can be expressed in terms of one-loop order renomalized quantities as
\begin{eqnarray}
\frac{ g_{a\bar{b}}}{\lambda_{0}} &=& \mu^{2\epsilon} \Bigl{\{} \frac{ g_{a\bar{b}}}{\lambda} + \frac{1}{2 \epsilon} \Bigl( - \frac{N}{2\pi} g_{a\bar{b}} \Bigr) + \cdots \Bigr{\}}   \nonumber \\
&=& \mu^{2\epsilon} \frac{ g_{a\bar{b}}}{\lambda} \Bigl( 1 - \frac{1}{2 \epsilon} \frac{\lambda N}{2\pi} \Bigr) ,  \label{g_0}
\end{eqnarray}
where $\mu$ is the scale parameter.
    \subsection{C-deformed sector}
We now evaluate the loop corections to the other coupling constant $C^{\alpha \beta}$. There are two graphs which give possibly divergent contributions to the C-deformed sector. They are shown in Fig.1. Other one-loop graphs contributing to the C-deformed sector are finite. In Fig.1 the external lines represent a set of background fields, such as 
\begin{eqnarray}
\delta_{a \bar{b}} \delta_{c\bar{d}} \epsilon^{\mu \nu} \partial_{\mu} \bar{\varphi}^{\bar{b}} \partial_{\nu}\bar{\varphi}^{\bar{d}}  C^{11} \chi^{a}_{+} \chi^{c}_{+}. \label{ex_BF}
\end{eqnarray}

\vspace{0.5cm}
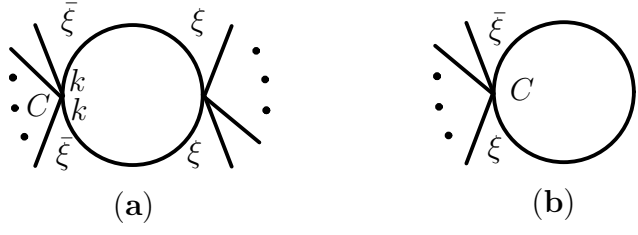
\begin{figure}[htbp]
\begin{center}

%WinTpicVersion3.08
\unitlength 0.1in
\begin{picture}( 32.6100,  9.8100)(  3.3000,-13.2100)
% CIRCLE 0 0 3 0
% 4 968 821 1258 1048 2354 803 4301 790
% 
\special{pn 20}%
\special{ar 968 822 368 368  6.2738846 6.2831853}%
\special{ar 968 822 368 368  0.0000000 6.2701990}%
% LINE 0 0 3 0
% 4 603 828 458 456 597 828 458 1193
% 
\special{pn 20}%
\special{pa 604 828}%
\special{pa 458 456}%
\special{fp}%
\special{pa 598 828}%
\special{pa 458 1194}%
\special{fp}%
% STR 2 0 3 0
% 3 410 867 410 930 2 0
% $C$
\put(4.1000,-9.3000){\makebox(0,0)[lb]{$C$}}%
% LINE 0 0 3 0
% 4 1340 831 1485 460 1346 831 1485 1197
% 
\special{pn 20}%
\special{pa 1340 832}%
\special{pa 1486 460}%
\special{fp}%
\special{pa 1346 832}%
\special{pa 1486 1198}%
\special{fp}%
% STR 2 0 3 0
% 3 641 888 641 951 2 0
% $k$
\put(6.4100,-9.5100){\makebox(0,0)[lb]{$k$}}%
% STR 2 0 3 0
% 3 565 1153 565 1216 2 0
% $\bar{\xi}$
\put(5.6500,-12.1600){\makebox(0,0)[lb]{$\bar{\xi}$}}%
% STR 2 0 3 0
% 3 590 447 590 510 2 0
% $\bar{\xi}$
\put(5.9000,-5.1000){\makebox(0,0)[lb]{$\bar{\xi}$}}%
% STR 2 0 3 0
% 3 1252 1140 1252 1203 2 0
% $\xi$
\put(12.5200,-12.0300){\makebox(0,0)[lb]{$\xi$}}%
% STR 2 0 3 0
% 3 1271 453 1271 516 2 0
% $\xi$
\put(12.7100,-5.1600){\makebox(0,0)[lb]{$\xi$}}%
% STR 2 0 3 0
% 3 634 737 634 800 2 0
% $k$
\put(6.3400,-8.0000){\makebox(0,0)[lb]{$k$}}%
% CIRCLE 0 0 3 0
% 4 3223 808 3513 1035 4609 789 6556 777
% 
\special{pn 20}%
\special{ar 3224 808 368 368  6.2738846 6.2831853}%
\special{ar 3224 808 368 368  0.0000000 6.2694777}%
% LINE 0 0 3 0
% 4 2858 815 2713 443 2852 815 2713 1180
% 
\special{pn 20}%
\special{pa 2858 816}%
\special{pa 2714 444}%
\special{fp}%
\special{pa 2852 816}%
\special{pa 2714 1180}%
\special{fp}%
% STR 2 0 3 0
% 3 2940 785 2940 848 2 0
% $C$
\put(29.4000,-8.4800){\makebox(0,0)[lb]{$C$}}%
% STR 2 0 3 0
% 3 2823 1118 2823 1181 2 0
% $\xi$
\put(28.2300,-11.8100){\makebox(0,0)[lb]{$\xi$}}%
% STR 2 0 3 0
% 3 2832 497 2832 560 2 0
% $\bar{\xi}$
\put(28.3200,-5.6000){\makebox(0,0)[lb]{$\bar{\xi}$}}%
% STR 2 0 3 0
% 3 863 1401 863 1491 2 0
% ({\bfseries a})
\put(8.6300,-14.9100){\makebox(0,0)[lb]{({\bfseries a})}}%
% STR 2 0 3 0
% 3 3055 1388 3055 1478 2 0
% ({\bfseries b})
\put(30.5500,-14.7800){\makebox(0,0)[lb]{({\bfseries b})}}%
% LINE 0 0 3 0
% 2 590 830 330 580
% 
\special{pn 20}%
\special{pa 590 830}%
\special{pa 330 580}%
\special{fp}%
% LINE 0 0 3 0
% 2 1340 830 1630 1070
% 
\special{pn 20}%
\special{pa 1340 830}%
\special{pa 1630 1070}%
\special{fp}%
% LINE 0 0 3 0
% 2 2850 813 2550 563
% 
\special{pn 20}%
\special{pa 2850 814}%
\special{pa 2550 564}%
\special{fp}%
% DOT 0 0 3 0
% 4 350 890 400 1040 340 730 340 730
% 
\special{pn 20}%
\special{sh 1}%
\special{ar 350 890 10 10 0  6.28318530717959E+0000}%
\special{sh 1}%
\special{ar 400 1040 10 10 0  6.28318530717959E+0000}%
\special{sh 1}%
\special{ar 340 730 10 10 0  6.28318530717959E+0000}%
\special{sh 1}%
\special{ar 340 730 10 10 0  6.28318530717959E+0000}%
% DOT 0 0 3 0
% 4 1660 740 1613 590 1667 901 1667 901
% 
\special{pn 20}%
\special{sh 1}%
\special{ar 1660 740 10 10 0  6.28318530717959E+0000}%
\special{sh 1}%
\special{ar 1614 590 10 10 0  6.28318530717959E+0000}%
\special{sh 1}%
\special{ar 1668 902 10 10 0  6.28318530717959E+0000}%
\special{sh 1}%
\special{ar 1668 902 10 10 0  6.28318530717959E+0000}%
% DOT 0 0 3 0
% 4 2570 883 2620 1033 2560 723 2560 723
% 
\special{pn 20}%
\special{sh 1}%
\special{ar 2570 884 10 10 0  6.28318530717959E+0000}%
\special{sh 1}%
\special{ar 2620 1034 10 10 0  6.28318530717959E+0000}%
\special{sh 1}%
\special{ar 2560 724 10 10 0  6.28318530717959E+0000}%
\special{sh 1}%
\special{ar 2560 724 10 10 0  6.28318530717959E+0000}%
\end{picture}%

\end{center}
\caption{ Divergent one-loop corections to the C-deformed sector. The solid internal lines stand for the $\xi$-propagator, the dots denote background-dependent vertices, such as (\ref{ex_BF}). }	
\label{}
\end{figure}
\vspace{0.5cm}

The loop graph 1.a involves two vertices, one containing $D_{\mu} \bar{\xi}^{\bar{b}} D_{\nu} \bar{\xi}^{\bar{d}}$ from $\mathcal{L}_{C}^{(2)}$ and another containing $\xi^{a} \xi^{c}$ from $\mathcal{L}_{{\rm Boson}}^{(2)}$.
\begin{eqnarray}
\mathcal{L}_{{\rm Boson}}^{(2)} &=& \cdots - \frac{1}{2} R_{a\bar{b}c\bar{d}} \partial_{\mu} \bar{\varphi}^{\bar{b}} \partial^{\mu} \bar{\varphi}^{\bar{d}} \xi^{a} \xi^{c} + \cdots   \label{Lag_B_2_0}  , \\
\mathcal{L}_{C}^{(2)} &=& \cdots + g_{a\bar{b}} g_{c\bar{d}} \epsilon^{\mu \nu} ( C^{11} \chi^{a}_{+} \chi^{c}_{+} - C^{22} \chi^{a}_{-} \chi^{c}_{-} ) D_{\mu} \bar{\xi}^{\bar{b}} D_{\nu} \bar{\xi}^{\bar{d}}  + \cdots  . \label{Lag_C_0_2}  
\end{eqnarray}
\vspace{0.3cm}
The graph 1.a contributes to the effective action the following term. 
\begin{eqnarray}
\delta \mathcal{L}_{C}^{(a)} & = & ( \cdots ) \times g_{a\bar{b}} g_{c\bar{d}} R_{i\bar{j}k\bar{l}} \partial^{\sigma} \bar{\varphi}^{\bar{j}} \partial^{\sigma} \bar{\varphi}^{\bar{l}} \epsilon^{\mu \nu} D_{\mu}  \dot{\bar{\xi}}^{\bar{b}} D_{\nu} \ddot{\bar{\xi}}^{\bar{d}} \times \dot{\xi}^{i} \ddot{\xi}^{k} \nonumber \\
&=& ( \cdots ) \times \epsilon_{\mu \nu} ( \quad I^{\mu \nu} \quad + \quad \mbox{finite} \quad ) ,   \label{div_1} 
\end{eqnarray}
where $I^{\mu \nu}$ is given by
\begin{eqnarray}
I^{\mu \nu} &=& \int_{0}^{1} dx \int d^2 k \frac{k^{\mu}k^{\nu}}{\{k^2 + M^2 (x)\}^2 } \nonumber \\ 
&& \nonumber \\
&\sim & \frac{1}{\epsilon} \delta^{\mu \nu}  .
\end{eqnarray}
Here $M^2 (x)$ does not contain $k^{\mu}$. We note that $I^{\mu \nu}$ is symmetric in $\mu$,$\nu$ and hence
\begin{eqnarray}
\delta \mathcal{L}_{C}^{(a)} & \sim & \epsilon_{\mu \nu} I^{\mu \nu} = 0.
\end{eqnarray}

The loop graph 1.b involves a single vertex, the one containing $\xi^{i}\bar{\xi}^{\bar{j}}$ from $\mathcal{L}_{C}^{(2)}$.
\begin{eqnarray}
\mathcal{L}_{C}^{(2)} &=& ( g_{a\bar{b}} R_{c\bar{d}i\bar{j}} + g_{c\bar{d}} R_{a\bar{b}i\bar{j}} ) \partial_{\mu}\bar{\varphi}^{\bar{b}} \partial_{\nu}\bar{\varphi}^{\bar{d}} \epsilon^{\mu \nu} ( C^{11} \chi^{a}_{+} \chi^{c}_{+} - C^{22} \chi^{a}_{-} \chi^{c}_{-} ) \xi^{i} \bar{\xi}^{\bar{j}}  + \cdots .  \label{Lag_C_1_1}  \nonumber \\
\end{eqnarray}
It's contribution to the effective action is computed to be 
\begin{eqnarray}
\delta \mathcal{L}_{C}^{(b)} & = & ( g_{a\bar{b}} R_{c\bar{d}i\bar{j}} + g_{c\bar{d}} R_{a\bar{b}i\bar{j}} ) \partial_{\mu}\bar{\varphi}^{\bar{b}} \partial_{\nu}\bar{\varphi}^{\bar{d}} \epsilon^{\mu \nu} ( C^{11} \chi^{a}_{+} \chi^{c}_{+} - C^{22} \chi^{a}_{-} \chi^{c}_{-} ) \dot{\xi}^{i} \dot{\bar{\xi}}^{\bar{j}}   \nonumber \\
                 & = & \frac{1}{2\epsilon} \frac{N}{2\pi} ( g_{a\bar{b}} g_{c\bar{d}} + g_{c\bar{d}} g_{a\bar{b}} ) \partial_{\mu}\bar{\varphi}^{\bar{b}} \partial_{\nu}\bar{\varphi}^{\bar{d}} \epsilon^{\mu \nu} ( C^{11} \chi^{a}_{+} \chi^{c}_{+} - C^{22} \chi^{a}_{-} \chi^{c}_{-} ) .
\end{eqnarray}
We thus have
\begin{eqnarray}
\delta \mathcal{L}_{C} & = & \frac{1}{\epsilon} \frac{\lambda N}{2\pi} \mathcal{L}_{C}  .
\end{eqnarray}

To summarize, at one-loop, only the loop graph of 1.b gives a divergent contribution to the C-deformed sector. The result is proportional to $\mathcal{L}_{C}$, and hence it can be eliminated by the counter term. It assures the renormalizability of the model at one-loop order.

    \subsection{UV divergences in the C-deformed sector and renormalization}

We decompose $C^{\alpha \beta}$ into renormalization part and constant part by writing
\begin{eqnarray}
C^{\alpha \beta} &=& \gamma \tilde{C}^{\alpha \beta} . 
\end{eqnarray}
The dimensionless coupling constant $\gamma$ receives renormalization and $\tilde{C}^{\alpha\beta}$ is set to constant, in the Lagrangian (\ref{LC}).
We introduce $T^{a \bar{b} c \bar{d}}$ as
\begin{eqnarray}
T^{a \bar{b} c \bar{d}} & \equiv & \partial_{\mu}\bar{\varphi}^{\bar{b}} \partial_{\nu}\bar{\varphi}^{\bar{d}} \epsilon^{\mu \nu} \{ \tilde{C}^{11} \chi^{a}_{+} \chi^{c}_{+} - \tilde{C}^{22} \chi^{a}_{-} \chi^{c}_{-}  \}  .
\end{eqnarray}

The counter term of $\mathcal{L}_{C}$ is $(\mathcal{L}_{C})_{0} - \mathcal{L}_{C}$. By using the renormalization result (\ref{g_0}), we have
\begin{eqnarray}
\mathcal{L}_{C,{\rm ct}} = (\mathcal{L}_{C})_{0} - \mathcal{L}_{C} &=& \lambda_{0} \gamma_{0} \frac{g_{a\bar{b}}}{\lambda} \frac{g_{c\bar{d}}}{\lambda} \Bigl( 1 - \frac{1}{2\epsilon} \frac{\lambda N}{2\pi} \Bigr)^2 T^{a \bar{b} c \bar{d}} - \lambda \gamma \frac{g_{a\bar{b}}}{\lambda} \frac{g_{c\bar{d}}}{\lambda} T^{a \bar{b} c \bar{d}}    \nonumber  \\
&=& \Bigl\{ \lambda_{0} \gamma_{0} \Bigl( 1 - \frac{1}{2\epsilon} 2 \frac{\lambda N}{2\pi} \Bigr)  - \lambda \gamma \Bigr\} \frac{g_{a\bar{b}}}{\lambda} \frac{g_{c\bar{d}}}{\lambda} T^{a \bar{b} c \bar{d}} \nonumber  \\
&=& \Bigl\{ \frac{\lambda_{0} \gamma_{0} }{\lambda \gamma} \Bigl( 1 - \frac{1}{2\epsilon} 2 \frac{\lambda N}{2\pi} \Bigr)  - 1 \Bigr\} \mathcal{L}_{C} . 
\end{eqnarray}
It is added to cancel the infinity from the loop graph, which we have computed above.
\begin{eqnarray}
\mathcal{L}_{C,{\rm ct}} + \delta \mathcal{L}_{C} &=& \Bigl( \frac{\lambda_{0} \gamma_{0} }{\lambda \gamma} - 1 \Bigr) \Bigl( 1 - \frac{1}{2\epsilon} 2 \frac{\lambda N}{2\pi} \Bigr) \mathcal{L}_{C} \nonumber  \\
&=& 0 .
\end{eqnarray}
Thus the renormalization of $\gamma$ is fixed by the condition
\begin{eqnarray}
\lambda_{0} \gamma_{0}  = \lambda \gamma . \label{reno_C}
\end{eqnarray}
We have found that the NC parameter $C^{\alpha \beta}$ (or $\gamma$) is renormalized at one-loop order as
\begin{eqnarray}
\gamma_0 / \gamma = (\lambda_0 / \lambda )^{-1} =  \mu^{2\epsilon} \Bigl( 1 - \frac{1}{2 \epsilon} \frac{\lambda N}{2\pi} \Bigr)  .
\end{eqnarray}

    \subsection{$\beta$-functions}
The ordinary 2D SUSY $CP^{N-1}$ model has good UV properties. We now study the UV properties of the deformed 2D SUSY $CP^{N-1}$ model, in particular its $\beta$-functions. The $\beta$-function of the coupling constant $\lambda$ is obtained from eq.(\ref{g_0}).
\begin{eqnarray}
\beta_{\lambda} &=& \mu \frac{\partial}{\partial \mu} \lambda = - \frac{N}{2\pi}\lambda^{2}.
\end{eqnarray}
For the purpose of perturbative computation we have defined the fields by setting
\begin{eqnarray}
\varphi^a , \chi^a & \to & \sqrt[]{\mathstrut \lambda} \varphi^a , \quad \sqrt[]{\mathstrut \lambda} \chi^a , \\
g_{a\bar{b}} &=& \frac{(1 + \lambda \bar{\varphi} \varphi ) \delta_{a\bar{b}} - \lambda \bar{\varphi}_{a} \varphi_{\bar{b}}}{(1 + \lambda \bar{\varphi}\varphi)^2} .
\end{eqnarray}
$\mathcal{L}_C$ is then written as
\begin{eqnarray}
\mathcal{L}_C &=& \lambda \gamma g_{a\bar{b}} g_{c\bar{d}} ( \tilde{C}^{11} \chi^{a}_{+} \chi^{c}_{+} - \tilde{C}^{22} \chi^{a}_{-} \chi^{c}_{-} ) \epsilon^{\mu \nu} (\partial_{\mu}\bar{\varphi}^{\bar{b}})(\partial_{\nu}\bar{\varphi}^{\bar{d}})  
.
\end{eqnarray}
We take $\lambda \gamma$ instead of $\gamma$ as the new coupling constant. The $\beta$-function of $\lambda \gamma$ vanishes.
\begin{eqnarray}
\beta_{\lambda \gamma} = \mu \frac{\partial}{\partial \mu} (\lambda \gamma) = 0 .
\end{eqnarray}
We note that $\lambda = 0$ is the UV fixed line of the theory in the $\lambda$ - $\gamma \lambda$ plane.

\section*{Acknowledgements} 
This work is supported partially by  the research grant of Japanese
Ministry of Education and and Science (Kiban B and Kiban C) and Chuo University research grant.
K.A. is supported by the Research Assistant Fellowship of Chuo University. 
H. N. is supported by Grant-in-Aid for Scientific Research 
from the Japanese Ministry of Education and Science, No. 16028203 for the priority area 
``origin of mass".

\end{document}